\documentstyle[11pt,twoside]{article}
\begin{document}

\title{Building dwarf galaxies out of merged young star clusters}
\author{Michael Fellhauer}
\date{Inst.\ f\"{u}r Theor.\ Phys.\ \& Astrophys.\ Univ.\ Kiel, 
	Leibnizstr.~15, 24098~Kiel, Germany}
\maketitle
\begin{abstract}
  Young star clusters in interacting galaxies are often found in
  groups or clusters of star clusters containing up to 100 single
  clusters.  In our project we study the future fate of these
  clusters of star clusters.  We find that the star clusters
  merge on time scales of a few dynamical crossing times of the
  super-cluster.  The resulting merger object has similarities
  with observed dwarf ellipticals (dE).  Furthermore, if
  destructive processes like tidal heating, dynamical friction or
  interaction with disc or bulge of the parent galaxy are taken
  into account our merger objects may evolve into objects
  resembling dwarf spheroidal galaxies (dSph), without the need
  of a high dark matter content. 
\end{abstract}

\section{Introduction}
\label{sec:intro}

Systems of interacting galaxies (like The Antennae or Stephan's
Quintet) show intense star formation.  In the tidal tails one
finds especially young massive star clusters ($R_{\rm eff}
\approx 2-10$~pc; $M \approx 10^{4}-10^{7}$~M$_{\odot}$; see
e.g. Whitmore et al. (1999)).  These star clusters are not
distributed uniformly but are clustered into groups, some
containing more than a hundred clusters.  With this project we
investigate the future fate of these clusters of star clusters
(what we call ``super-cluster'').  Following the arguments of
Kroupa (1998) these super-clusters may be bound entities, meaning
that the star clusters in the super-cluster are not only bound to
the tidal field of the parent galaxy but also bound to each
other.  Therefore, the star clusters are very likely to merge
with each other on a few dynamical times of the super-cluster.
This merging process and the resulting merger object are the
major topic of this project.   

\section{Setup}
\label{sec:setup}

We simulate super-clusters containing $N_{0}=20$ star
clusters.  Each of them is represented by a Plummer sphere,
containing 100,000 particles, a total mass of
$10^{6}$~M$_{\odot}$, a scale-length (Plummer radius) of $R_{\rm
  pl}=6$~pc and a crossing time of 1.4~Myr.  All star clusters
are distributed into an encompassing Plummer sphere with varied
scale-length, $R_{\rm pl}^{\rm sc}= 50, 75, 150, 300$~pc, and this
super-cluster is placed on a circular orbit at distance $D= 5,
10, 20, 30, 50, 100$~kpc around an analytic galactic potential
with an asymptotically flat rotation curve with $v_{\rm c} =
220$~km/s and $R_{\rm gal} = 4$~kpc.  For the discussion of the
results we define two dimensionless parameters for our systems.
$\alpha = R_{\rm pl} / R_{\rm pl}^{\rm sc} = 0.02, 0.04, 0.08,
0.12$ is the ratio between the scale-length of the single cluster
and the scale-length of the super-cluster.  $\beta = R_{\rm
  cut}^{\rm sc} / R_{\rm tidal} = 0.2$--$2.7$ is the ratio
between the size and the tidal radius of the super-cluster.
Several random realizations of each parameter combination are
performed to reduce stochastic errors.  The simulations are
carried out with the particle mesh code {\sc Superbox} (Fellhauer
et al. 2000) which has the ability to trace many objects with
different levels of high resolution sub-grids.  While the
outermost grid covers the whole orbit around the galactic centre
with the lowest resolution, the medium sized grids of each star
cluster cover the size of the super-cluster and all intermediate
interactions between the clusters in medium resolution
(10--50~pc), and the innermost high resolution (1~pc) grids cover
each single cluster.  Both medium and high resolution grids stay
focused on the simulated objects and travel with them through the
simulation area.  

\section{Compact Merger Objects}
\label{sec:dE}

A detailed description of the merging time-scales and the
efficiency of the merger process will appear in Fellhauer et al.\
(2001). 

The resulting merger objects are compact, spherical and dense
objects.  They have half-mass radii from 40--150 pc and sizes 
of 120--1,000 pc.  The mass of these objects resembles between
75 and 95~\% of the total mass of the initial super-cluster.
Central densities are derived by fitting exponential profiles to
the inner parts.  The range of central densities is
100--1,000~M$_{\odot}$/pc$^{3}$ with exponential scale lengths of
about 10~pc.  In the outer parts a steep power law ($r^{-3.5}$)
fits the density best.  The surface-density can be well fitted
with a de-Vaucouleur profile.  The velocity dispersion in the
central region is about 30~km/s but measured along the
line-of-sight through the centre one derives about 20~km/s.  All
this data suggests that our merger objects have similar
properties to nucleated dwarf galaxies (dE,N).  Because the
merger process lasts for about a dozen crossing times of the
super-cluster (for small $\alpha$'s this can sum up to more than 
1~Gyr) the merger object is surrounded by several star clusters
which are still in the process of merging or are on near-circular
orbits around the super-cluster and slowly decaying through
dynamical friction.  Therefore, our model can account for high
specific cluster frequencies ($S_{\rm N}$), which are found for
dE,N-galaxies.

Furthermore, our simple models do not account for destructive
processes like tidal heating on eccentric orbits, dynamical
friction in the parent galaxy and decay of orbit, shocks due to
disc or bulge passages.  All these processes would lead to a
mass-loss of the merger object and a decrease in its central
density.  Translated to observations, our objects would transform
into dE,noN with time, while the number of remaining clusters is
also decreasing.  Observations show that dE,N have higher $S_{\rm
  N}$ than dE,noN.

Our objects also show extended tidal tails along the whole orbit
around the galactic centre.  Escaped star clusters are found in
these tidal features.

\section{Dissolved Merger Object}
\label{sec:dSph}

In two test calculations the super-cluster was set up on an
eccentric orbit ($D_{\rm apo} = 60$~kpc, $D_{\rm peri} = 30$~kpc)
so that tidal heating could act on the merger objects.  The
merger objects get dissolved with time (mass-loss $\approx 5$~\%
per perigalacticon passage).  Due to a disruptive interaction
with a second object the central densities drop down from several
hundred to 0.3--0.002~M$_{\odot}$/pc$^{3}$, but can still be
fitted with an exponential in the innermost part.  But in the
outer parts the material is not bound to the object anymore.  It
has a ``fluffy'' structure extending a few kpc.  The
surface-density can not be fitted with a de-Vaucouleur profile
anymore.  The measured central velocity dispersions are in the
order of 1--3~km/s, but measured along the line-of-sight through
the centre and the outer unbound parts one derives about
10~km/s.  Compared with observations these objects look like
dwarf spheroidal galaxies (c.f. Klessen \& Kroupa 1998).  Star
clusters remaining around these objects can account for the very
high $S_{\rm N}$ of some dSph-galaxies. 

\section{Conclusions}
\label{sec:conc}

With our simulations we are able to show, that the star clusters
in a super-cluster merge efficiently.  Comparing our parameters
with observations we predict an exponential decrease with time of
the number of clusters if the observed super-clusters are bound
entities. 

The resulting merger object has similar properties like
dE,N-galaxies, and if destructive processes are taken into account
the next step in the evolution of our objects would be dE,noN.
This also gives a natural explanation why dE,N have higher $S_{\rm
  N}$ than dE,noN.

The final stage of the evolution would be a dSph-galaxy,
suggesting that at least some of these objects may not have a
predicted high dark matter content. \\

{\bf References:}\\

\noindent Fellhauer, M., Kroupa, P., Baumgardt, H., Bien, R., 	
Boily, C.M., Spurzem, R., \& Wassmer, N. 2000,  NewA, {\bf 5}, 305 \\
Fellhauer, M., \& Kroupa P. 2000, ASP Conf.\ Ser., {\bf 211}, 241$^{1}$ \\
Fellhauer, M., Baumgardt, H., Kroupa, P., \& Spurzem, R. 2001, 
Cel.\ Mech.\ \& Dyn.\ Astron. in press; astro-ph/0103052 \\
Gallagher, S.C., Hunsberger, S.D., Charlton, J.C., \& Zaritsky, D. 
2000, ASP Conf.\ Ser., {\bf 211}, 247$^{1}$ \\
Grebel, E.K. 2000, ASP Conf.\ Ser., {\bf 211}, 262$^{1}$ \\ 
Hunsberger, S.D., Gallagher, S.D., Charlton, J.C., \& Zaritsky, D. 
2000, ASP Conf. Ser., {\bf 211}, 254$^{1}$ \\
Klessen, R.S., \& Kroupa, P. 1998, ApJ, {\bf 498}, 143 \\ 
Kroupa, P. 1998, MNRAS, {\bf 300}, 200 \\
Mateo, M. 1998, ARAA, {\bf 36}, 435 \\
Whitmore, B.C., Zhang, Q., Leitherer, C., \& Fall, S.M. 1999, 
AJ, {\bf 118}, 1551 \\
Zhang, Q., \& Fall, S.M. 1999,  ApJL, {\bf 527}, 81L \\

\noindent $^{1}$: ASP Conf.\ Ser.\ 211: Proceedings of the workshop
`Massive Stellar Clusters' held in Strasbourg Nov.\ 1999; eds.\
A. Lancon, C.M. Boily

\end{document}